\documentclass{PoS}

\usepackage{amsmath,amssymb,amsfonts}
\usepackage[english]{babel}
\usepackage{graphicx,psfrag}
\usepackage{epsfig}
\usepackage{epstopdf}

\usepackage{slashed}

\newcommand{\ev}[1]{\left < #1 \right >}

\newcommand{\be}{\begin{equation}}
\newcommand{\ee}{\end{equation}}
\newcommand{\bea}{\begin{eqnarray}} 
\newcommand{\eea}{\end{eqnarray}}

\newcommand{\bmp}{\noindent\begin{minipage}{16cm}}

\usepackage{cite}

\title{Wilson Fermions with Four Fermion Interactions}

\ShortTitle{ Four Fermion Interactions with Wilson Fermions}

\author{\speaker{Jarno Rantaharju}\\
       CP3 -Origins, IFK \& IMADA, University of Southern Denmark\\
       E-mail: \email{rantaharju@cp3.dias.sdu.dk}}

\author{Vincent Drach\\
        CP3 -Origins, IFK \& IMADA, University of Southern Denmark\\
        E-mail: \email{drach@cp3.dias.sdu.dk}}

\author{Ari Hietanen\\
        CP3 -Origins, IFK \& IMADA, University of Southern Denmark\\
        E-mail: \email{hietanen@cp3.dias.sdu.dk}}
        
\author{Claudio Pica\\
       CP3 -Origins, IFK \& IMADA, University of Southern Denmark\\
       E-mail: \email{pica@cp3.dias.sdu.dk}}
       
\author{Francesco Sannino\\
       CP3 -Origins and the Danish IAS,  University of Southern Denmark\\
       E-mail: \email{sannino@cp3.dias.sdu.dk}}

\abstract{We present a lattice study of a four fermion theory, known as Nambu Jona-Lasinio (NJL) theory, via Wilson fermions. Four fermion interactions naturally occur in several  extensions of the Standard Model as a low energy parameterisation of a more fundamental theory.  In models of dynamical electroweak symmetry breaking these operators, at an effective level, are used to endow the Standard Model fermions with masses.  Furthermore these operators, when sufficiently strong, can drastically modify the fundamental composite dynamics by, for example, turning a strongly coupled infrared conformal theory into a (near) conformal one with desirable features for model building.  As first step, we study  spontaneous chiral symmetry breaking for the lattice version of the NJL model. 
\vskip 0.2cm
{\noindent  Preprint: CP$^3$-Origins-2015-045 DNRF90, DIAS-2015-45}
}

\FullConference{The 33rd International Symposium on Lattice Field Theory\\
		 14-18 July, 2015\\
		 Kobe International Conference Center, Kobe, Japan}

\begin{document}

\section{Introduction}

Despite the apparent Standard Model-like nature of the Higgs sector, it might still conceal, in plain sight, a more fundamental composite nature \cite{Krog:2015bca}.  In fact, it has been recently shown that gauge-Yukawa theories, similar to the Standard Model, even if perturbative for some energy range can still abide compositeness conditions \cite{Bardeen:1989ds,Krog:2015bca}. In this case the Higgs-like state at higher energies is not a propagating degree of freedom and effective four-fermion interactions appear, and the theory becomes a gauged NJL theory.   

Traditional models of fundamental composite dynamics, of either Technicolor or composite Goldstone Higgs, make also extensive use of four-fermion interactions.  This is clear from a  recent attempt  \cite{Cacciapaglia:2015yra} to generate the top mass in a composite theory able to bridge between the Technicolor and Composite Goldstone Higgs limit.
  
For these kind of models three types of four-fermion operators generally occur:
\begin{align*}
&L_\text{eff} = \frac{a}{\Lambda^2_{UV} } (\bar\Psi_{SM}\Psi_{SM})^2  + \frac{b}{\Lambda^2_{UV} } \bar\Psi_{SM}\Psi_{SM}\bar\Psi_{TC} \Psi_{TC}  + \frac{c}{\Lambda^2_{UV} } (\bar\Psi_{TC}\Psi_{TC})^2.
\end{align*}
The first term, involving only Standard Model fermions, can be suppressed by the large cutoff scale $\Lambda^2_{UV}$. The other two may be enhanced by the dynamics of the Technicolor sector.  The second term provides the Standard Model fermion masses and the third one adds to the Technicolor dynamics. 

According to Holdom's insight \cite{Holdom:1981rm},   models of walking dynamics, possessing  large mass anomalous dimensions, can enhance the SM fermion mass-term operator dynamically. It was later recognised \cite{Fukano:2010yv} that an ideal way to achieve walking dynamics (iWalk) is to ask the third operator (i.e. the one containing only the technifermions) to induce chiral symmetry breaking when added to an otherwise conformal Technicolor dynamics \cite{Yamawaki:1996vr,Fukano:2010yv}. Our ultimate goal is therefore to investigate the nonperturbative dynamics of the gauged NJL model. We investigate on the lattice the un-gauged version, i.e. the pure Nambu Jona-Lasinio (NJL) model, with Wilson fermions and furthermore we retain only the last four-fermion operator.  This is the stepping stone for the gauged version.  
  A similar model has been previously studied with the goal of understanding the phase structure of Wilson fermions \cite{Bitar:1993cs,Bitar:1993xi,Aoki:1993vs}.  Models with staggered fermions have been studied in previous works \cite{Hasenfratz:1991it,AliKhan:1993dx,Kogut:1998rg,Sinclair:1998ji,Catterall:2011ab} and chiral symmetry breaking has been observed.

It is worth noting that the model is only effective. It is not renormalizable and does not have a continuum limit. Thus different lattice discretizations and the continuum are essentially separate models, although they share fundamental properties. There is, however, a scaling region where physical results depend only weakly on the lattice spacing $a$. 
%

\section{The Model}

We study the NJL model with 2 flavors of fermions possessing 2 extra {\it color} degrees of freedom. In this case one can construct an NJL action that preserves the full chiral symmetry. Unfortunately when representing the fermion fields with pseudofermions, the action must be rendered quadratic using auxiliary fields and the fermion determinant becomes complex\footnote{It is possible to render the fermion determinant positive if the number of colors is even and there is no gauge interaction \cite{Bitar:1993cs,Bitar:1993xi,Aoki:1993vs}. The remedy is not applicable here since we plan to generalize the study to a gauged model. }. We will therefore study a model that preserves only a $U(1)\times U(1)$ subgroup of the original $SU(2)\times SU(2)$ symmetry group.

The model is defined by the continuum Lagrangian
\begin{align}
& \tilde L = \bar\Psi \slashed \partial \Psi + \sigma \bar\Psi\Psi+ \pi \bar\Psi i \gamma_5\tau_3 \Psi +\frac{\sigma^2+\pi^2}{4\gamma^2} \label{cont_tL} \\
 &\ev{\sigma} = 2\gamma^2 \,\ev{\bar\Psi\Psi} \,\,\,\,\,\,
 \ev{\pi} = 2\gamma^2 \ev{\bar\Psi  i\gamma_5 \tau_3 \Psi}. \label{aux_exps}
\end{align}
Using the Wilson discretization of the pseudofermion representation we have:
\begin{align}
& L = \chi^\dagger (M^{\dagger}M)^{-1} \chi +\frac{\sigma^2+\pi^2}{4\gamma^2} \label{lat_L} \\
& M=\slashed \partial_W  + \sigma + \pi i \gamma_5
\end{align}

The Wilson term breaks  chiral symmetry and introduces additional explicit chiral symmetry breaking terms in the Lagrangian. In a renormalizable model the only necessary correction, a mass term, arises as a result of a divergence in the Wilson term \cite{Bochicchio:1985xa}. It is manifestly of  first order in $a$. All other terms vanish at small $a$. In the model under question new terms may also arise with powers of $\gamma$ and diverge at small $a$.

Chiral symmetry is restored by a correction to the action such that the axial current $\left ( A_\mu^3 \right )$ is conserved:
\begin{align*}
\partial_\mu \ev{A^3_\mu(x)  O} = 0
\end{align*}
The correction may be found, for example,  via the PCAC relation. With the action \ref{lat_L} we have
\begin{align}
\partial_\mu \ev{A^3_\mu(x)O} = \ev{ aX^3(x) O }, \label{PCACeq}
\end{align}
where $aX^3$ is the variation of the Wilson term. This term includes the divergent contribution that is cancelled by the fermion mass term. It is instructive write down it's renormalisation, expanding to the first few lowest order operators:
\begin{align}
 aX^3(x) &= a\bar X^3(x) + \frac{c_m\left( \gamma/a \right)}{a} P^3(x) + \left ( Z_A\left( \gamma/a \right) -1 \right ) \partial_\mu A^3_\mu(x) \label{Xexpansion} \\
  &+ a c_{A,1}\left( \gamma/a \right) \partial_\mu \partial_\mu P^3(x) + a^2 c_{A,2}\left( \gamma/a \right) \partial_{(\mu} \partial_\nu \partial_{\nu)} A_\mu^3(x) \nonumber \\
 &+ a^2 c_\gamma\left( \gamma/a \right)  S^0(x)P^3(x) + \cdots  \nonumber
 \label{pcac} 
\end{align}
where $a\bar X^3(x)$ includes only vanishing contributions to the PCAC relation, $P^3$ is the pseudoscalar density and $S^0$ is the scalar density. The second term on the right hand side has to be cancelled by adding a mass term to the action and the third one is a multiplicative correction to the axial current. The two terms on the second line can be seen as corrections to the axial current. The last term includes the first new contribution to the action with an effect on chiral symmetry. It is related to choosing the quadratic coupling of the auxiliary fields to be different.

Each coefficient in eq. \ref{Xexpansion} can be expanded in $\gamma$ to reveal contributions divergent in $a$. This is a natural consequence of the nonrenormalizable nature of the model. Any number of these terms may be taken into account in the action and in the renormalization of lattice operators. For simplicity we study a model with only the mass correction. We include the $c_\gamma$ term only to study its effect.
The full Lagrangian is thus
\begin{align}
& L = \chi^\dagger (M^{\dagger}M)^{-1} \chi +\frac{\sigma^2}{4\gamma^2} + \frac{\pi^2}{4\gamma^2+\delta_\gamma}\\
& M=\slashed \partial_W  + m_0+ \sigma + \pi i \gamma_5. 
\end{align}
In the following we refer to the physical mass as $m=m_0+c_m/a$ and the correction as $\bar \delta = \delta_\gamma + a^2c_\gamma$. The restoration of chiral symmetry implies that all the terms on the right hand side in equation \ref{PCACeq} cancel.

At a given $\gamma$, when $a$ is too small, the divergent contributions may become relevant. On the other hand, when $a$ is too large, the corrections arising with positive order of $a$ will be significant. Only the region in between will provide physically interesting results.
\\

An additional complication arises because of the auxiliary field $\pi(x)$. There are significant disconnected contributions in the mesonic triplet channels.
The disconnected part increases with $\gamma$, and is relevant for the pseudoscalar correlators.

The meson masses are measured from correlators of the type
\begin{align}
& C_{\small \Gamma}(t_0,t) = \ev{ \sum_{\bf x}\bar\Psi({\bf x}, t_0) \Gamma \tau^3 \Psi({\bf x}, t_0)  \sum_{\bf y}\bar\Psi({\bf y}, t) \Gamma \tau^3 \Psi({\bf y}, t)   }\\
& \lim_{t-t_0 \rightarrow \infty} C_{\small \Gamma}(t_0,t) = A_{\small \Gamma} e^{ m_{\small \Gamma} (t_0-t) }
\end{align}
For the vector meson mass ($m_\rho$) it is sufficient to measure the correlator with $\Gamma=\gamma_k$ and fitting it to the exponential at large $t-t_0$. This correlator does not suffer significantly from disconnected contributions. We measure the pseudoscalar mass ($m_\pi$) by from the correlator with  $\Gamma = \gamma_0\gamma_5$ and, when possible,  $\Gamma = \gamma_5$ using the generalized eigenvalue method. We also measure the pseudoscalar mass ($m_{\pi_2}$) using the correlator 
\begin{align}
& C_\pi(t_0,t) = \ev{ \sum_{\bf x} \pi({\bf x}, t_0) \pi({\bf y}, t)   }
\end{align}
This measurement is useful when the disconnected contribution is large. The correlator is noisy, but can be evaluated without inverting the fermion matrix.

\section{Numerical Results}

First we study the correction $\bar\delta$ and its effect on chiral symmetry and the fermion mass. To this end we simulate the model with a small coupling $\gamma = 0.2a$ and several values of $\delta_\gamma$. In all cases the lattice size is $16\times 8^3$. When $\bar \delta$ is nonzero, the chiral symmetry is not restored at zero fermion mass $m$. We can therefore study the effect of the correction by separately finding the line where $m=0$ and the chirally symmetric line.

To find the chirally symmetric line we measure the pion mass $m_\pi$ and the divergence of the axial current normalized by the pseudoscalar density
\begin{align}
\tilde m = \frac{ \sum_{{\bf x},{\bf y }} \partial_{x_0} \ev{A_\mu^3(x_0,{\bf x}) A^3_\mu(0,{\bf y}) }}{ \sum_{{\bf x},{\bf y }} \ev{ P^3(x_0,{\bf x}) A^3_\mu(0,{\bf y})} }.
\end{align}
The restoration of chiral symmetry implies $\tilde m =0$.

To study the mass term we must define it without reference to the axial current. To determine a zero mass value we use $m_\sigma = m_0+\ev{\sigma}$. At small coupling this coincides with the fermion mass. In general the expectation value $\ev{\sigma}$ is related to the chiral condensate (equation \ref{aux_exps}). At a finite lattice size there is always a small but nonzero chiral condensate, which changes sign when the fermion mass crosses zero. This is seen as a first order transition in $m_\sigma$. A similar method was used in meanfield theory in \cite{Izubuchi:1998hy}.

\begin{figure}
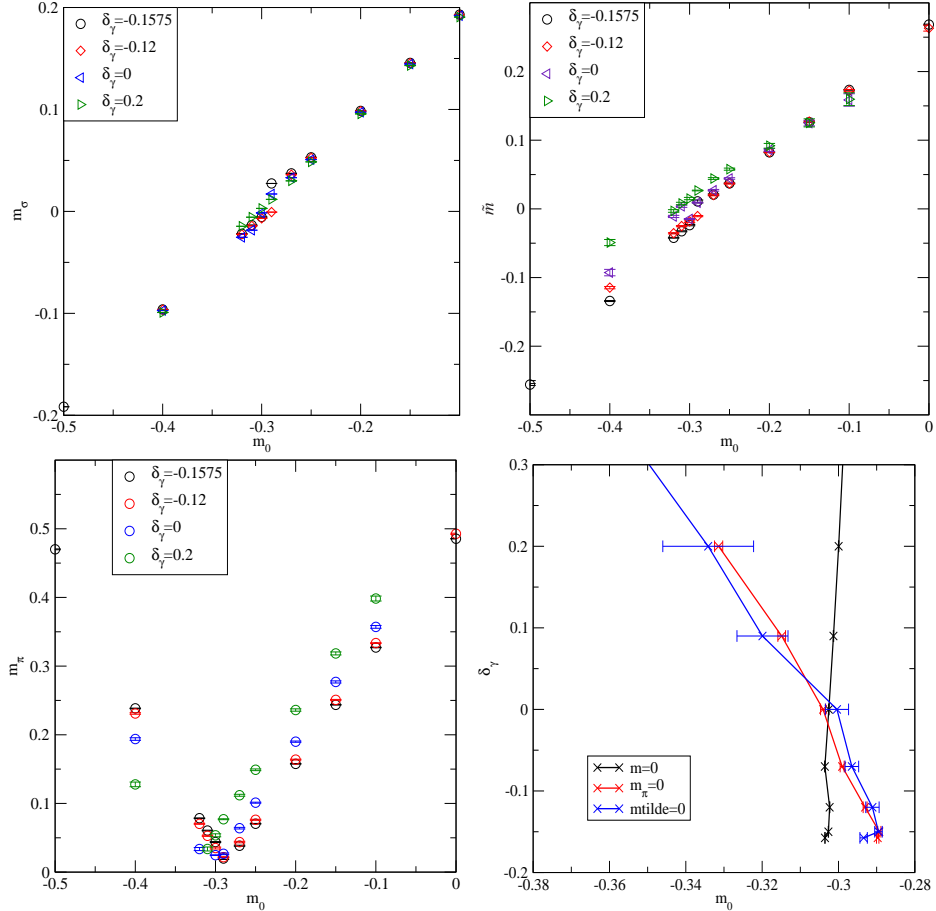
 \center
\includegraphics[width=.4\linewidth,height=.4\linewidth]{msigma.eps}
\psfrag{mtilde}[][][0.6]{$\tilde m$}
{
    \includegraphics[width=.4\linewidth,height=.4\linewidth]{mtilde.eps}
}\\
\includegraphics[width=.4\linewidth,height=.4\linewidth]{mpi.eps}
\psfrag{mtilde}[][][0.5]{$\tilde m$} {
\includegraphics[width=.4\linewidth,height=.4\linewidth]{phaseplot.eps}
}
\caption{ Fermion and meson masses at small coupling.We have  $\gamma =  0.2a$ for several values of $\delta_\gamma$. Upper left: the bare propagator fermion mass. Upper right: $\tilde m$. Lower left: the pion mass. Lower right: the points in the $\delta_\gamma$ and $m_0$ plane where the each mass crosses zero. }
\label{small_coupling_masses}
\end{figure}

\begin{figure}
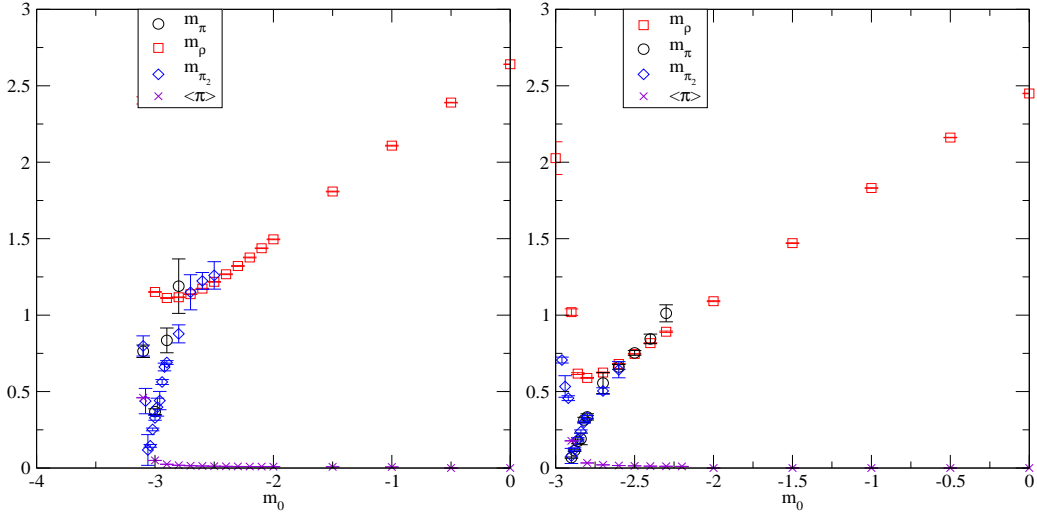
 \center
\includegraphics[width=.45\linewidth,height=.45\linewidth]{{mpirho1.3_L8}.eps}
\includegraphics[width=.45\linewidth,height=.45\linewidth]{{mpirho1.2_L8}.eps}
\caption{ The pseudoscalar and vector meson masses large coupling $0.65a$ (left) and $0.6a$ (right). We also show the expectation value $\ev{\pi}$, which serves as the order parameter for the parity broken phase.}
\label{large_coupling_masses}
\end{figure}

The masses are shown in figure \ref{small_coupling_masses}. We observe a jump in the $m_\sigma$ across zero when $\delta_\gamma \le 0$. At $\delta_\gamma>0$ we do not observe a jump. In each case we find $m_\sigma=0$ with a linear extrapolation using points where $\left | m_\sigma \right | > 0.05$.
We find the chirally symmetric point with linear extrapolations of $\tilde m$ and $m_\pi$ using points where $m_\pi>0.1$. The resulting values are shown in figure \ref{large_coupling_masses}.

When $\delta_\gamma=0$ the jump in $m_\sigma$ and the restoration of chiral symmetry happen at the same bare mass. This confirms the expectation that $a^2c_\gamma$ is negligible at small $\gamma$. At $\delta_\gamma<0$  chiral symmetry is restored at  positive $m_\sigma$.  At $\delta_\gamma>0$ we do not observe a jump when $m_\sigma$ crosses zero, but chiral symmetry is restored at a negatife value of negative $m_\sigma$. In all cases a nonzero value for the correction produces a noticeable but relatively small difference between the two critical lines.

Having established the restoration of chiral symmetry at small coupling, we study its spontaneous breaking at large coupling. We run simulations at $\gamma=0.65a$ and $0.6a$ and $\delta_\gamma=0$. The lattice size is again $16\times8^3$. We measure the pseudoscalar meson masses $m_\pi$ and $m_{\pi_2}$ and the vector meson mass $m_\rho$. Chiral symmetry is restored when $m_\pi=m_{\pi_2}=0$ and broken when  $m_\rho$ is different from $m_\pi$ and $m_{\pi_2}$.

We show the masses of the pseudoscalar and the vector mesons in figure \ref{large_coupling_masses}. We also show the expectation value $\ev{\pi}$, the order parameter for the parity broken phase. At large mass the disconnected contributions to the pseudoscalar correlators become problematic. It is clear, however, that at small mass the chiral symmetry is broken. In both cases the mass of the vector state remains finite through the transition to the parity broken phase, where as $m_\pi$ and $m_{\pi_2}$ approach zero at the transition.

\section{Conclusions}

We presented a lattice study of an NJLmodel, via Wilson fermions. These studies are relevant because  four fermion interactions  occur naturally in several  extensions of the Standard Model. They represent low energy parameterisations of a more fundamental theory. Additionally models of dynamical electroweak symmetry breaking make extensive use of four-fermion operators  to endow the Standard Model fermions with masses.    We investigated spontaneous chiral symmetry breaking for the lattice version of the NJL model presented here.

One observes that disconnected contributions to the correlator of the pseudoscalar meson become significant at large coupling. Evaluating the disconnected contribution requires significantly more computational effort than evaluating the connected part. At relatively small mass and large coupling the correlator of the auxiliary field $\pi$ can be used instead. 

According to the expectations we have find evidence of spontaneous chiral symmetry breaking at  large coupling, $\gamma=0.65a$ and $0.6a$. The results are compatible with previous studies and  follows meanfield results \cite{Bitar:1993cs,Bitar:1993xi,Aoki:1993vs,Catterall:2011ab}. Additional work is needed to cover the full phase space of the model and to verify scaling and lattice size dependence. Once these analyses are performed it is then natural to move to the full gauged model.

\section{Acknowledgments}

We thank Michele Della Morte for useful discussion. This work was supported by the Danish National Research Foundation DNRF:90 grant and by a Lundbeck Foundation Fellowship grant. The computing facilities were provided by the Danish Centre for Scientific Computing.

\end{document}